\documentclass[sigconf,nonacm]{acmart}
\usepackage{subfigure}

\usepackage{amsmath,amsfonts}
\usepackage{algorithmic}
\usepackage{graphicx}
\usepackage{textcomp}
\usepackage{xcolor}
\usepackage{cases}
\usepackage{multirow}
\usepackage{booktabs}
\usepackage[capitalize]{cleveref}
\usepackage{makecell}
\usepackage{color, soul}
\usepackage{bbding}
\usepackage{colortbl}
\usepackage{arydshln}
\usepackage[group-separator={,}]{siunitx}
\usepackage[most]{tcolorbox}
\newtcolorbox{mybox}[1][]{
breakable,
  arc=1mm,
  boxrule=1pt,
  colback=yellow!14,
  colframe=black!80,
  fonttitle=\bfseries,
  title=#1,
  left=1mm,
  right=1mm,
  top=1mm,
  bottom=1mm
}
\AtBeginDocument{%
  \providecommand\BibTeX{{%
    \normalfont B\kern-0.5em{\scshape i\kern-0.25em b}\kern-0.8em\TeX}}}

\setcopyright{acmcopyright}
\copyrightyear{2024}
\acmYear{2024}
\acmDOI{XXXXXXX.XXXXXXX}

\acmConference[ICPC 2024]{46th International Conference on Program Comprehension}{April 2024}{Lisbon, Portugal}
\acmPrice{15.00}
\acmISBN{978-1-4503-XXXX-X/18/06}

\begin{document}

\title{Interpretable Online Log Analysis Using Large Language Models with Prompt Strategies}

\author{Yilun Liu$^1$, Shimin Tao$^{1,2}$, Weibin Meng$^1$, Jingyu Wang$^3$, Wenbing Ma$^1$,\\ Yuhang Chen$^2$, Yanqing Zhao$^1$, Hao Yang$^1$, Yanfei Jiang$^1$} 
\affiliation{\vspace{0.05cm}\institution{$^1$ Huawei\country{China}}
\institution{$^2$ University of Science and Technology of China\country{China}}
\institution{$^3$ Beijing University of Posts and Telecommunications\country{China}}
}
\email{{liuyilun3,taoshimin, mengweibin3, mawenbing, zhaoyanqing, yanghao30,jiangyanfei}@huawei.com}
\email{wangjingyu@bupt.edu.cn, chen97@mail.ustc.edu.cn}
\renewcommand{\shortauthors}{Yilun Liu, Shimin Tao, and Weibin Meng, et al.}
\begin{abstract}
Automated log analysis is crucial in modern software-intensive systems for facilitating program comprehension throughout software maintenance and engineering life cycles. Existing methods perform tasks such as log parsing and log anomaly detection by providing a single prediction value without interpretation. However, given the increasing volume of system events, the limited interpretability of analysis results hinders analysts' comprehension of program status and their ability to take appropriate actions. Moreover, these methods require substantial in-domain training data, and their performance declines sharply (by up to 62.5\%) in online scenarios involving unseen logs from new domains, a common occurrence due to rapid software updates. In this paper, we propose LogPrompt, a novel interpretable log analysis approach for online scenarios. LogPrompt employs large language models (LLMs) to perform online log analysis tasks via a suite of advanced prompt strategies tailored for log tasks, which enhances LLMs' performance by up to 380.7\% compared with simple prompts. Experiments on nine publicly available evaluation datasets across two tasks demonstrate that LogPrompt, despite requiring no in-domain training, outperforms existing approaches trained on thousands of logs by up to 55.9\%. We also conduct a human evaluation of LogPrompt's interpretability, with six practitioners possessing over 10 years of experience, who highly rated the generated content in terms of usefulness and readability (averagely 4.42/5). LogPrompt also exhibits remarkable compatibility with open-source and smaller-scale LLMs, making it flexible for practical deployment. Code of LogPrompt is available at \url{https://github.com/lunyiliu/LogPrompt}.
\end{abstract}



\begin{teaserfigure}
\centering
 \subfigbottomskip=-2pt 
 \subfigcapskip=-2pt 
 \subfigure[Existing workflow of automated log analysis]{
  \includegraphics[width=0.48\linewidth]{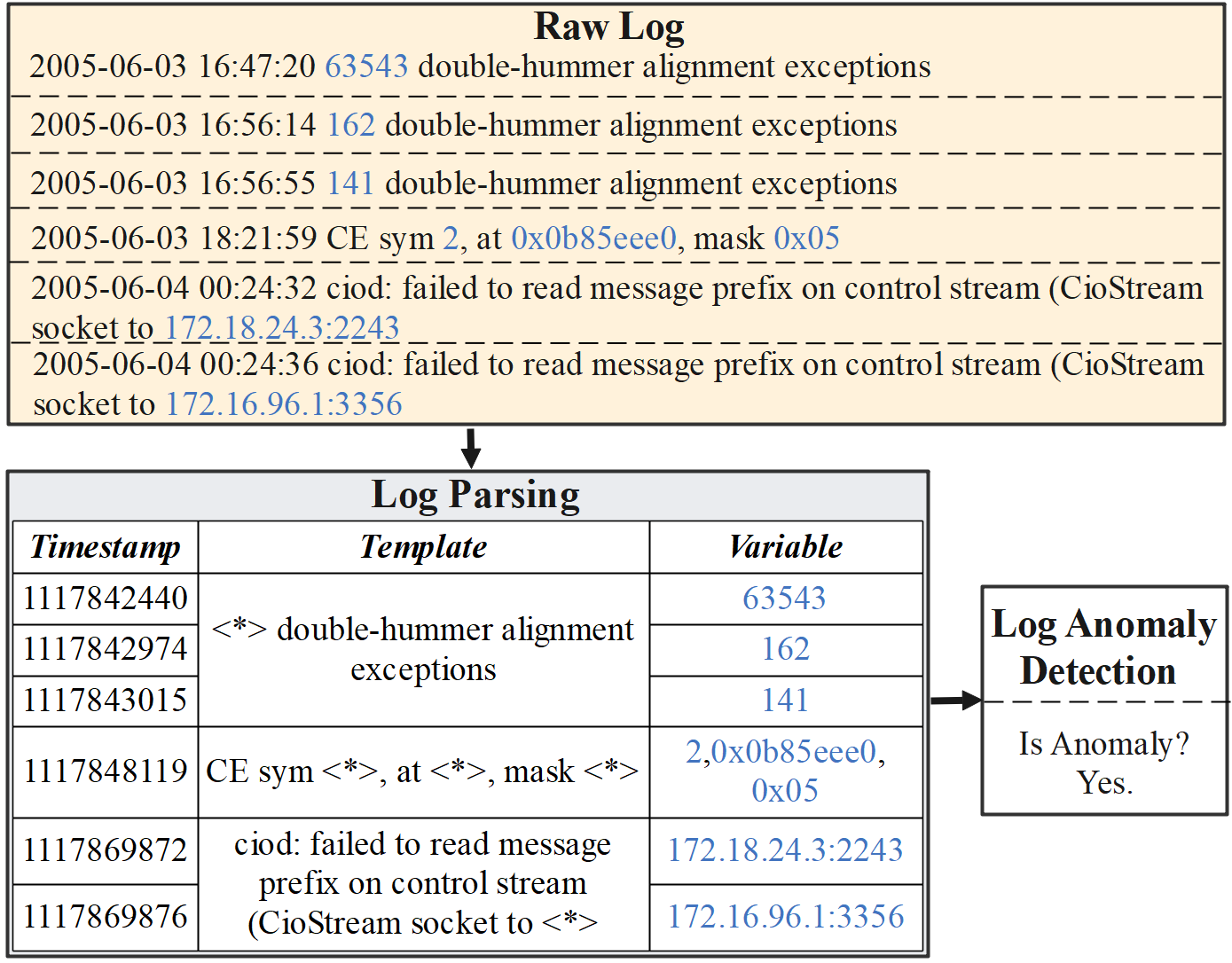}}
 \subfigure[Interpretable log analysis]{
  \includegraphics[width=0.493\linewidth]{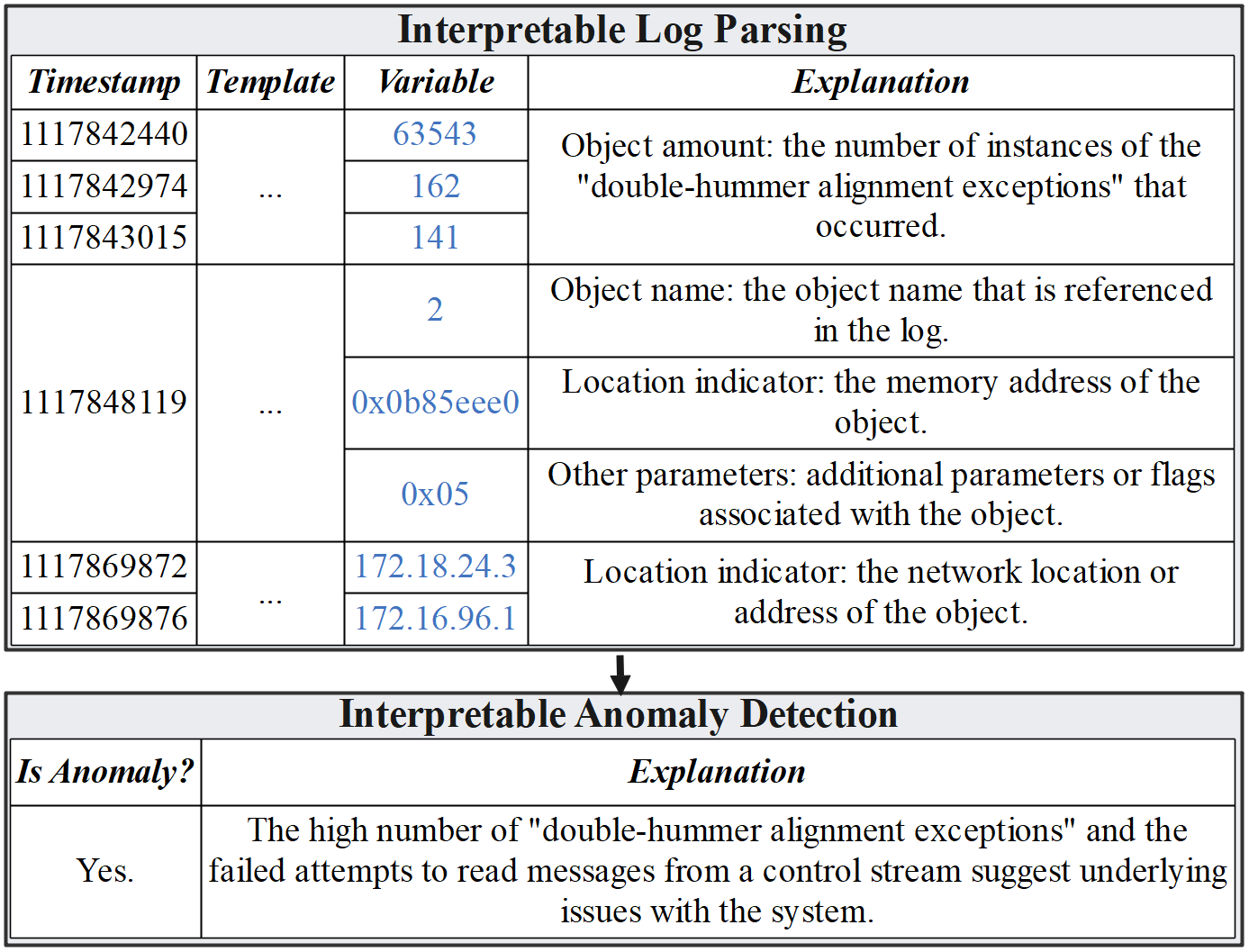}}
 \caption{Illustration of automated log analysis: (a) without interpretability and (b) with interpretability. The presented explanations are generated by our approach on 2023/04/18.}
\label{LogAnalysis}
\end{teaserfigure}


\maketitle

\section{Introduction}

Automated log analysis is essential for ensuring the reliability and resilience of fundamental services in the development and maintenance of software-intensive systems by facilitating the comprehension of program status. For instance, a popular e-commerce website serving millions of global users daily may operate on a cluster of hundreds or thousands of servers, generating vast amounts of log data that records system failures, performance degradation, and unexpected interruptions. Traditionally, when a failure occurs in a large-scale system, Operations and Maintenance (O\&M) engineers must first review copious system logs, identify abnormal events, determine root causes, and then implement appropriate mitigation measures. This process is time-consuming, requires multiple rounds of communication among engineers, and demands advanced expertise. In contrast, current automated log analysis techniques can provide efficient approaches for program comprehension, such as log abstractions (through log parsing) and anomaly detection, significantly reducing the burden on O\&M engineers. As illustrated in Fig. \ref{LogAnalysis}(a), log parsing extracts common (static) segments ($template$) and unique (dynamic) segments ($variable$) from raw logs, minimizing log size while retaining key information; and log anomaly detection identifies anomalies in historical log sequences.

Though existing approaches have achieved remarkable performance in log analysis when combined with deep learning techniques \cite{le2022log,tao2022logstamp,li2023did}, challenges arise when log analysis technologies are applied in real-world scenarios. First, the performance of current methods significantly declines when training samples are scarce and unseen logs are the majority, a situation common in the industry (referred as the online scenario \cite{meng2020logparse,tao2022logstamp,biglog}). In online scenarios, operators frequently perform software upgrades on services to introduce new features, resolve bugs, or enhance performance, resulting in the generation of new types of logs and the potential incompatibility of logs from older versions. Consequently, the availability of an insufficient number of historical logs often hinders effective model training. Thus, there are usually not enough historical logs to effectively train the model. In extreme cases, when a brand new service goes online, no training samples are available due to the lack of in-domain logs, necessitating an highly adaptive log analysis approach that can be applied to new domains without in-domain training. 

Second, the potential benefits of log analysis techniques in improving program comprehension are hampered by their limited interpretability. Traditional methods, as depicted in Fig. \ref{LogAnalysis}(a), provide only prediction without further explanation. This absence of justification makes it less easier in some cases for human analysts to comprehend and act on analysis results, while an interpretable analysis output, as shown in Fig. \ref{LogAnalysis}(b), not only aids program comprehension with ample explanations and justifications, but also facilitates detecting false alarms, tracing root causes and taking appropriate actions.

In this paper, we aim to address two relatively unexplored challenges in log analysis, namely online scenarios and limited interpretability, by leveraging the recent advancements in large language models (LLMs). Our findings indicate that enhancing prompt strategies is crucial in unleashing the log analysis capabilities of LLMs and tackling these challenges. To this end, we propose a novel log analysis approach called LogPrompt, which encompasses three prompt strategies: (1) self-prompt, (2) chain-of-thought (CoT) prompt, and (3) in-context prompt. Experiments on nine public evaluation datasets across log parsing and log anomaly detection tasks reveal that LogPrompt not only achieves SOTA performance without in-domain training process, but also generates useful justifications for its predictions to enhance the interpretability of log analysis. These generated justifications are rated by six senior practitioners with over 10 years of experience in software and system O\&M. Our contributions can be summarized as follows:
\begin{itemize}
    \item We present a highly adaptive online log analysis approach that requires nearly no in-domain training. While most existing approaches require training on thousands of in-domain logs, our approach outperforms them by up to 55.9\% without requiring substantial in-domain training.
    \item We take the first step towards achieving interpretability in log analysis. In the human evaluation with six senior practitioners, over 80\% of the automatically generated explanations for variables and anomalies in log analysis are considered highly useful and readable, which could aid O\&M engineers in swiftly comprehending program status, composing analysis reports and ruling out false alarms.
    \item We propose a collection of prompt strategies designed for log analysis tasks, which consistently improve LLMs' performance by up to 380.7\% compared to simple prompts. Furthermore, the compatibility of these strategies with open-source and smaller-scale LLMs suggests their potential for use in industrial applications.

\end{itemize}
Additionally, we release the code of LogPrompt, which not only encourages further exploration of advanced prompting strategies when LLMs are employed in log analysis within the research community but also facilitates actual deployments by engineers.
\label{sec:introduction}

\section{Background}\label{sec:background}
\subsection{Large Language Models}
Large language models (LLMs) are trained using a probabilistic approach to generate text based on diverse training corpora. More specifically, these models focus on estimating the probability of the next word given a preceding sequence of words, employing an auto-regressive process whereby tokens are predicted one at a time until a designated end token is encountered~\cite{radford2019language}. The recent advancement of LLMs has brought a significant impact across multiple domains. For instance, ChatGPT~\cite{ouyang2022training}, a particularly powerful LLM, not only excels at various Natural Language Processing (NLP) tasks~\cite{peng2023towards,kocmi2023large,kung2023performance}, but also provides natural language explanations to justify its decision-making process. The activation of LLMs' capabilities relies on prompts, which typically take the form of textual templates with blank slots for LLMs to fill in \cite{liu2023pre}. Studies have demonstrated that advanced prompt strategies can significantly enhance the performance of LLMs~\cite{zhou2022large,white2023prompt,wei2022chain}.

\subsection{Pre-training \& Prompting}
\begin{figure}[htbp]
    \centering
  \includegraphics[width=\linewidth]{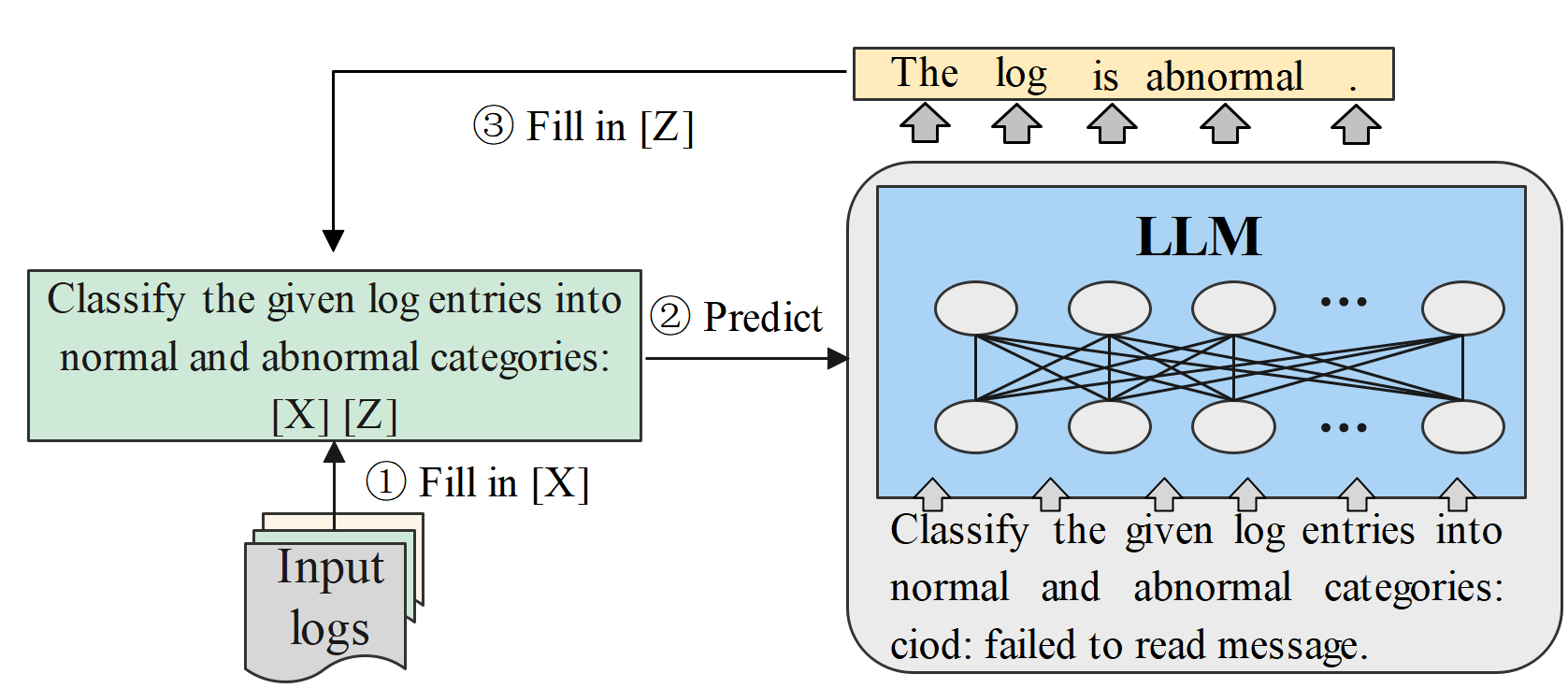}
  \caption{Illustration of utilizing a simple prompt for log anomaly detection with a large language model. \texttt{[X]} represents the input slot, while \texttt{[Z]} denotes the answer slot.}
  \label{fig2}
\end{figure}

The introduction of BERT \cite{kenton2019bert} heralded a pre-training \& fine-tuning paradigm within the NLP community. Such language models are pre-trained to extract robust, general-purpose language features and subsequently adapted to various downstream tasks by introducing additional parameters and fine-tuning them using task-specific objective functions. However, with the substantial increase in language model scale (e.g., from 220M parameters in BERT to 175B in ChatGPT) and their enhanced capabilities, this paradigm has evolved towards pre-training \& prompting \cite{liu2023pre}, due to the expensive cost of fine-tuning entire models. As depicted in Fig. \ref{fig2}, this approach involves designing a task-specific prompt to guide the LLM in generating the desired output, which is directly input to the model without altering its parameters.

A textual prompt can be viewed as a template comprised of three components: (1) the prompt prefix, a fixed string conveying task-specific information (\emph{e.g.}, the prompt prefix for the simple prompt in Fig. \ref{fig2} for anomaly detection is \texttt{Classify the given log entries into normal and abnormal categories:}); (2) the input slot \texttt{[X]}, which will later incorporate one or multiple task inputs (\emph{e.g.}, log entries); and (3) the answer slot \texttt{[Z]}. Once the input slot is populated with actual input, the only dynamic component of the prompt is \texttt{[Z]}, typically positioned at the end of the prompt. The LLM then predicts answers in an auto-regressive manner, populating the \texttt{[Z]} slot accordingly (\emph{e.g.}, with a binary word choice between ``normal'' and ``abnormal'').

\section{Design of LogPrompt}
\label{sec:design}
\subsection{Motivation}\label{sec:motivation}

\textbf{(1) Insufficient interpretability.} The goal of program comprehension is to facilitate engineers in understanding the program status, thereby reducing the cost for addressing breakdowns. However, as highlighted by recent studies~\cite{chen2021experience,huo2023semparser}, existing approaches only provide simple predictions without sufficient justification and semantics, thereby impeding engineers in comprehending the analysis results easily. For instance, in Fig. \ref{LogAnalysis}(a), traditional methods report only the positions of variables in log parsing and binary Yes/No responses for anomaly detection, which still require efforts from practitioners to interpret and verify analysis results, especially for unfamiliar logs in online scenarios. In contrast, Fig. \ref{LogAnalysis}(b) demonstrates the advantages of providing explanations along with parsed variables in logs, facilitating the interpretation of program status (\emph{e.g.}, the variable ``141'' is correctly interpreted by our approach as the number of occurrences of the ``double-hummer exceptions''). The justifications for detected anomalies also help comprehend the conclusion. For example, the explanation along with the anomaly conclusion in Fig. \ref{LogAnalysis}(b) further interprets the anomaly in the system, attributing it to the high number of ``double-hummer exceptions'' and the failed read attempt from a control stream, thereby validating the credibility of the prediction.
\begin{figure}[htbp]
 \centering  
 \subfigbottomskip=-2pt 
 \subfigcapskip=-2pt 
 \subfigure[Anomaly detection with ample/scarce training data]{
  \includegraphics[width=0.96\linewidth]{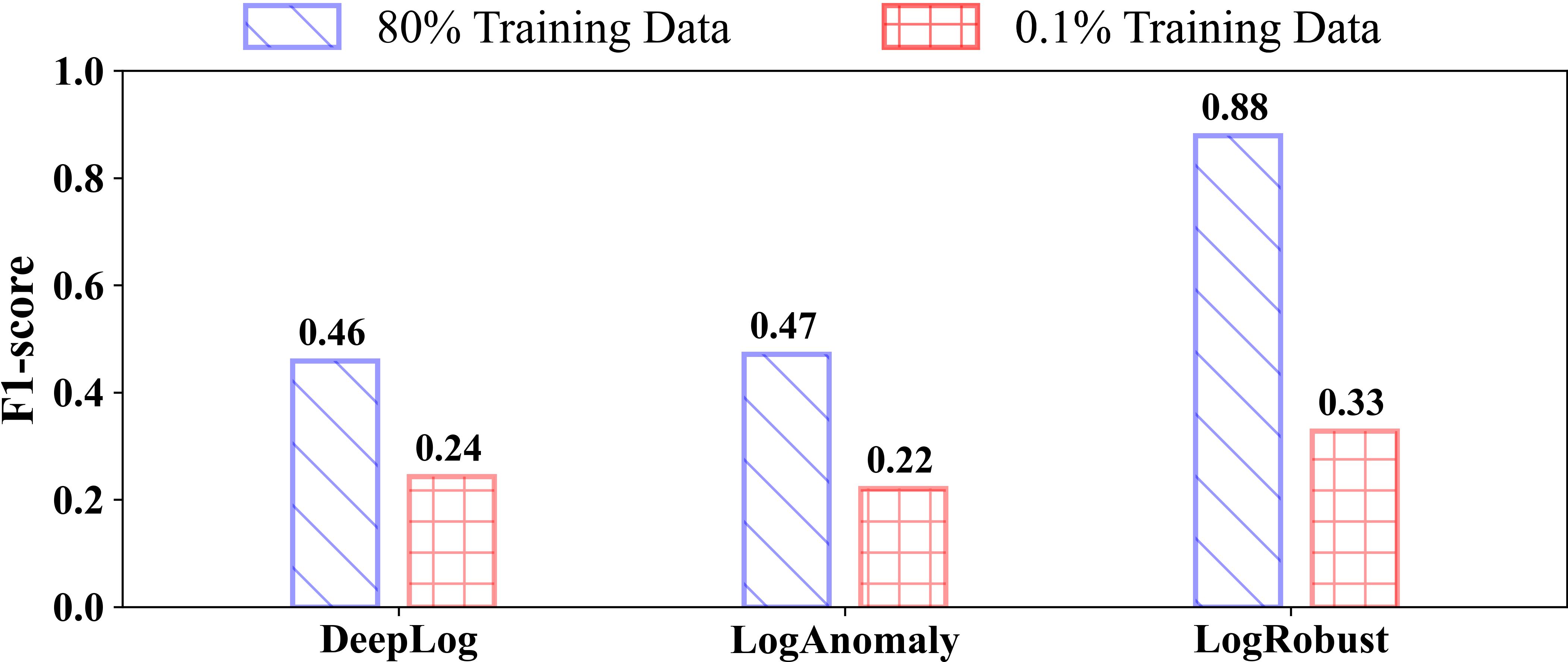}}
   \\
 \subfigure[Log parsing with ample/scarce training data]{
  \includegraphics[width=0.96\linewidth]{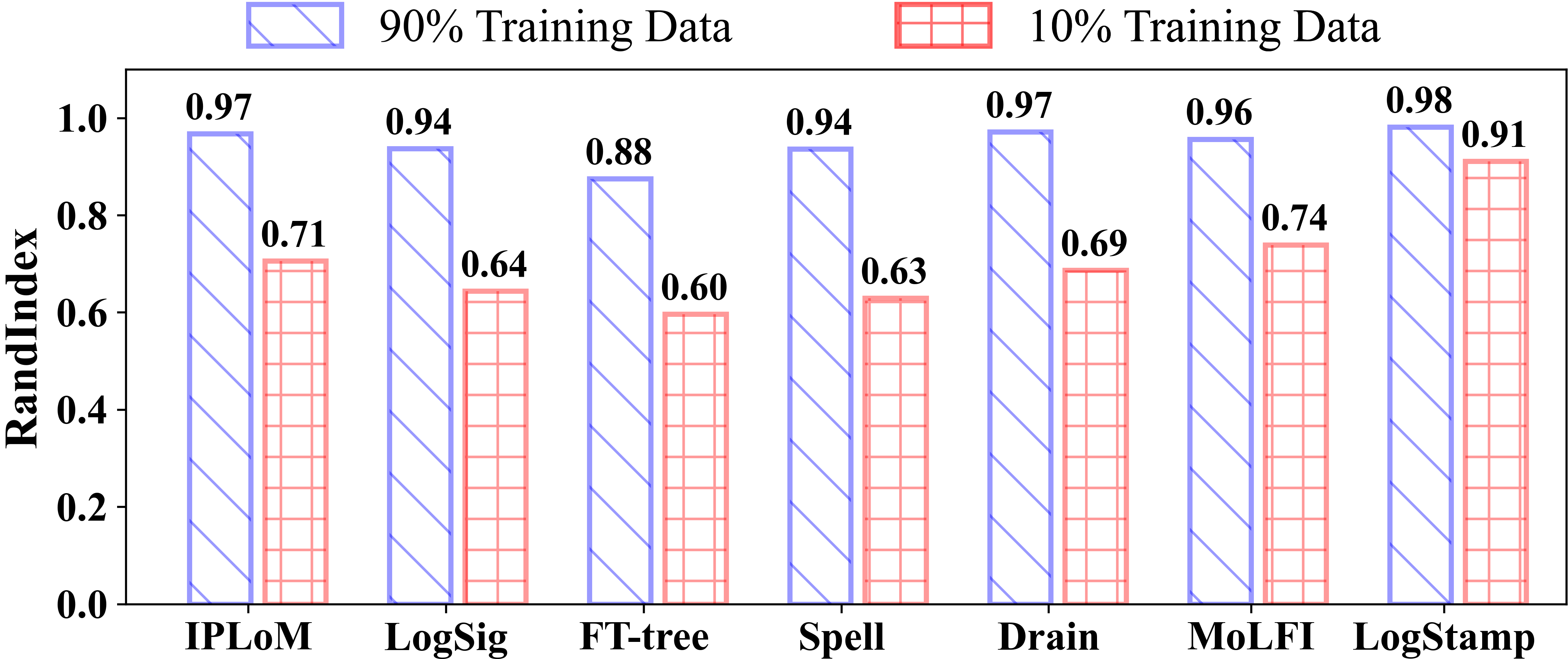}}
 \caption{The average performance of existing log analysis approaches deteriorates when in-domain logs for training are limited in availability.}

\label{fig3}
\end{figure}

\textbf{(2) Weak adaptability caused by data inefficiency.} In modern software systems, the generation of massive logs from devices across various domains has become increasingly common \cite{da-parser}. These logs are highly heterogeneous in format and terminology, and are frequently updated with new log events in the online scenario. To analyze such logs, an automated log analysis approach with strong adaptability is essential to ensure steady performance when being faced with new log events. However, previous studies \cite{tao2022logstamp} and our preliminary experiments have shown that existing approaches perform well only when ample historical logs with correct human labels are given for training, and their performance deteriorates drastically when in-domain training data is scarce. Fig.~\ref{fig3} presents the mean performance of existing approaches in the two log analysis tasks, averaged across two and seven public datasets for each respective task. When the ratio of training data (chronologically splitting~\cite{biglog,tao2022logstamp}) is 80\% or 90\% (tested on the remaining 20\% or 10\% logs in each dataset), existing approaches achieve nearly 100\% accuracy on both analysis tasks. However, when the available in-domain log data for training is limited to 10\% or less, and unseen logs are present in the test set, their performance drops drastically (\emph{e.g.}, the F1-score for LogRobust drops from 0.88 to 0.33). To effectively handle new log events, existing approaches require continuous training with newly labelled logs, and some may even need to be retrained from scratch. Therefore, an log analysis approach that can directly analyze unseen logs from any domain without the need for manually labeling massive historical logs and expensive training processes is desirable for the online scenario. 

\textbf{(3) Prompt strategies matter.} Despite the promising potential of leveraging LLMs such as ChatGPT for interpretable online log analysis due to their powerful language understanding abilities, directly applying a simple prompt to LLMs can result in poor performance. In our preliminary experiments, ChatGPT with the simple prompt shown in Figure \ref{fig2} achieved an F1-score of only 0.189 in anomaly detection. However, our best prompt strategy outperformed the simple prompt by 0.195 in F1-score. Similarly, in log parsing, we observed a performance fluctuation of up to 380.7\% among different prompt strategies. Recent studies have also reported performance differences between simple and advanced prompts when LLMs are used for NLP tasks \cite{jiao2023chatgpt,peng2023towards,kocmi2023large}. An unregulated prompt can lead to inaccurate results and even produce invalid content (\emph{e.g.}, hallucinations). Therefore, designing a task-specific prompt strategy is critical to effectively leverage LLMs to address various challenges in log analysis tasks with minimal efforts.

\subsection{Overview of LogPrompt}
The primary objective of LogPrompt is to enhance the correctness and interpretability of log analysis in the online scenario, through properly prompting LLMs. Firstly, LogPrompt standardizes the formats of input logs, by setting the context for slot \texttt{[X]}, and the formats of output answers (\texttt{[Z]}), in order to mitigate uncertainty in the generated solutions. Subsequently, a tailored prompt strategy is employed to construct the prompt prefix. The assembly of the prompt is completed by concatenating the prompt prefix, input slot, and answer slot, with the final response acquired from the LLM, following the procedure illustrated in Fig. \ref{fig2}.

We conducted experiments using three distinct strategies for constructing the prompt prefix, each founded on unique underlying principles: (1) \textit{self-prompt}, which aims to leverage the intrinsic capabilities of the LLM to generate pertinent prompt candidates for log analysis tasks; (2) \textit{chain-of-thought (CoT) prompt}, which emphasizes a systematic, step-by-step reasoning process by compelling the LLM to address log analysis problems through a sequence of intermediate steps, both explicitly and implicitly; (3) \textit{in-context prompt}, which supplies multiple log examples to establish a contextual understanding for the specific task.

\subsection{Format Controls}
To mitigate transmission failures due to internet fluctuations when invoking LLMs, such as ChatGPT, via an API, we insert multiple input logs into the \texttt{[X]} slot of a prompt until the length limitation for a single query is reached. However, our preliminary experiments revealed that prompts without format control statements often resulted in omitted logs, inconsistent formats, and unrelated content in the query results. Consequently, LogPrompt employs the following two functions, $f_x(\texttt{[X]})$ and $f_z(\texttt{[Z]})$, to establish the context for the input slot \texttt{[X]} and the answer slot \texttt{[Z]}, respectively.
\subsubsection{Input Control Function $f_x(\texttt{[X]})$}
Given $N$ input logs, $f_x$ constructs a context for the input slots $\texttt{[X]}_\texttt{1}$ \ldots $\texttt{[X]}_\texttt{N}$.
\begin{mybox}
$f_x(\texttt{[X]})=$ There are N logs, the logs begin: (1)$\texttt{[X]}_\texttt{1}$\verb|\n| (2)$\texttt{[X]}_\texttt{2}$\verb|\n| \ldots \verb|\n| (N)$\texttt{[X]}_\texttt{N}$
\end{mybox}
\subsubsection{Answer Control Function $f_z(\texttt{[Z]},S)$}
The answer control function adds prefix texts to \texttt{[Z]}, regulating the output answer format for each input log in a query. $S$ is a text string describing the desired answer value range. For instance, in anomaly detection, $S$ might be ``a binary choice between abnormal and normal'', while in log parsing, it could be ``a parsed log template''.
\begin{mybox}
$f_z(\texttt{[Z]},S)=$ Organize your answer to be the following format: (1) $x$-$y$\verb|\n| (2) $x$-$y$\verb|\n| \ldots \verb|\n| (N) $x$-$y$, where $x$ is $S$ and $y$ is the reason. \texttt{[Z]}
\end{mybox}

\subsection{Prompt Strategies}\label{sec:prompt_strategy}
\begin{figure}[htbp]
    \centering
  \includegraphics[width=\linewidth]{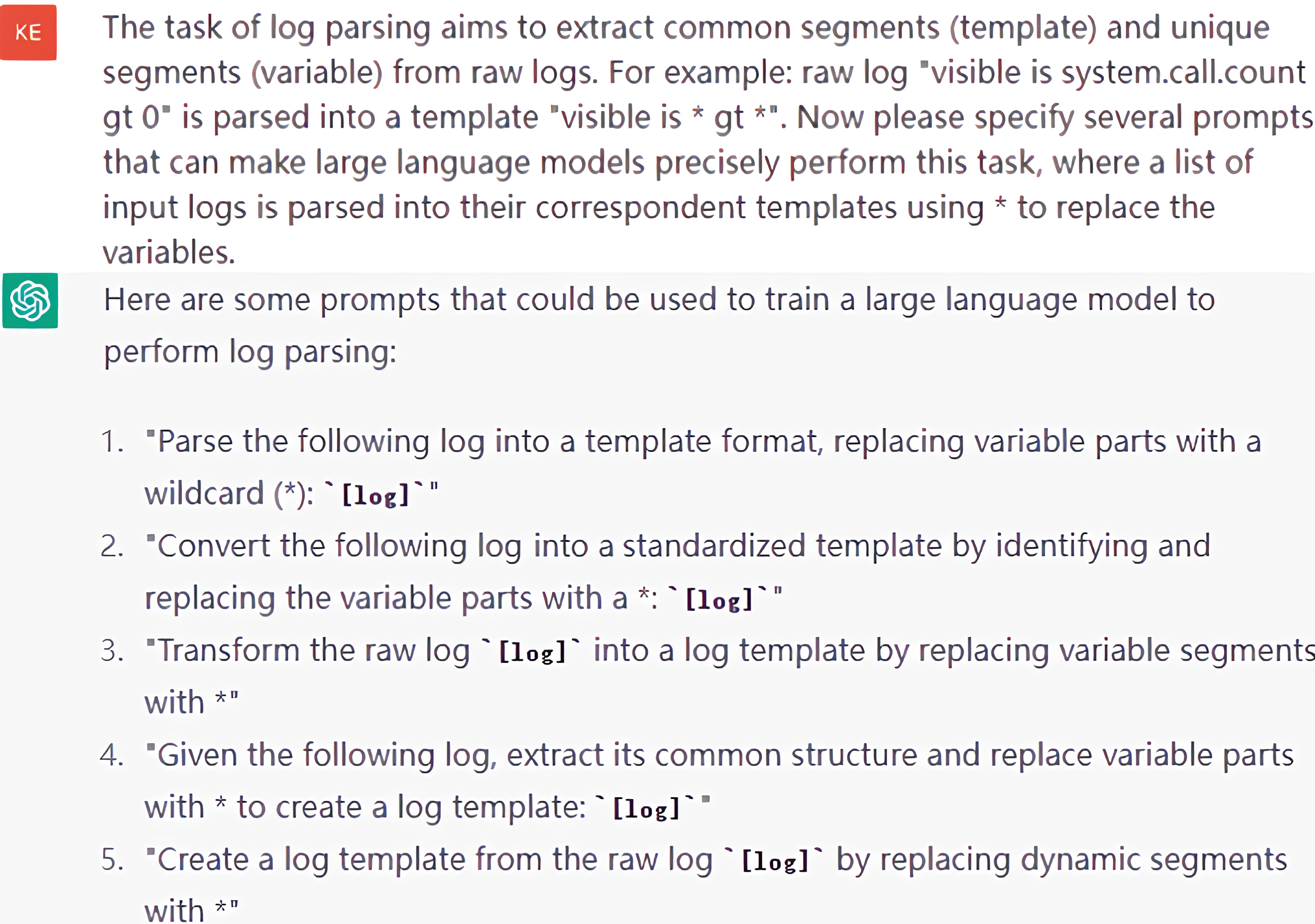}
  \caption{Prompts advised by ChatGPT for the task of log parsing (Dialogue date: 2023/02/25).}
  \label{fig4}
\end{figure}

\subsubsection{Self-prompt}\label{sec:self_prompt} Drawing inspiration from the work of Jiao \emph{et al.}\cite{jiao2023chatgpt}, we seek to create prompts that activate ChatGPT's log analysis capabilities by soliciting advice directly from ChatGPT itself. We prompt the LLM to generate several prompts that can guide LLMs in performing a specific task, as demonstrated in the log parsing task in Fig. \ref{fig4}. First, the self-prompt strategy prompts the LLM to generate a pool of prompt candidates $C$ using a prompt that consists of a precise task description and output requirements. The pool $C$ contains multiple prompts that ChatGPT believes can be used for log analysis tasks. Next, we evaluate the prompts in $C$ on a small task-specific log dataset, selecting the best prompt based on a metric score $s(p)$, where $p\in C$. Finally, with $S_p=$ ``a parsed log template'' representing the answer range string, the complete self-prompt for log parsing, combined with format control functions, can be expressed as:
\begin{mybox}
$Self$-$prompt$: $\arg\max_{p \in C} \{ s(p) \}\ f_x(\texttt{[X]})\ f_z(\texttt{[Z]},S_p)$ 
\end{mybox}
\begin{figure}[htbp]
    \centering
  \includegraphics[width=0.9\linewidth]{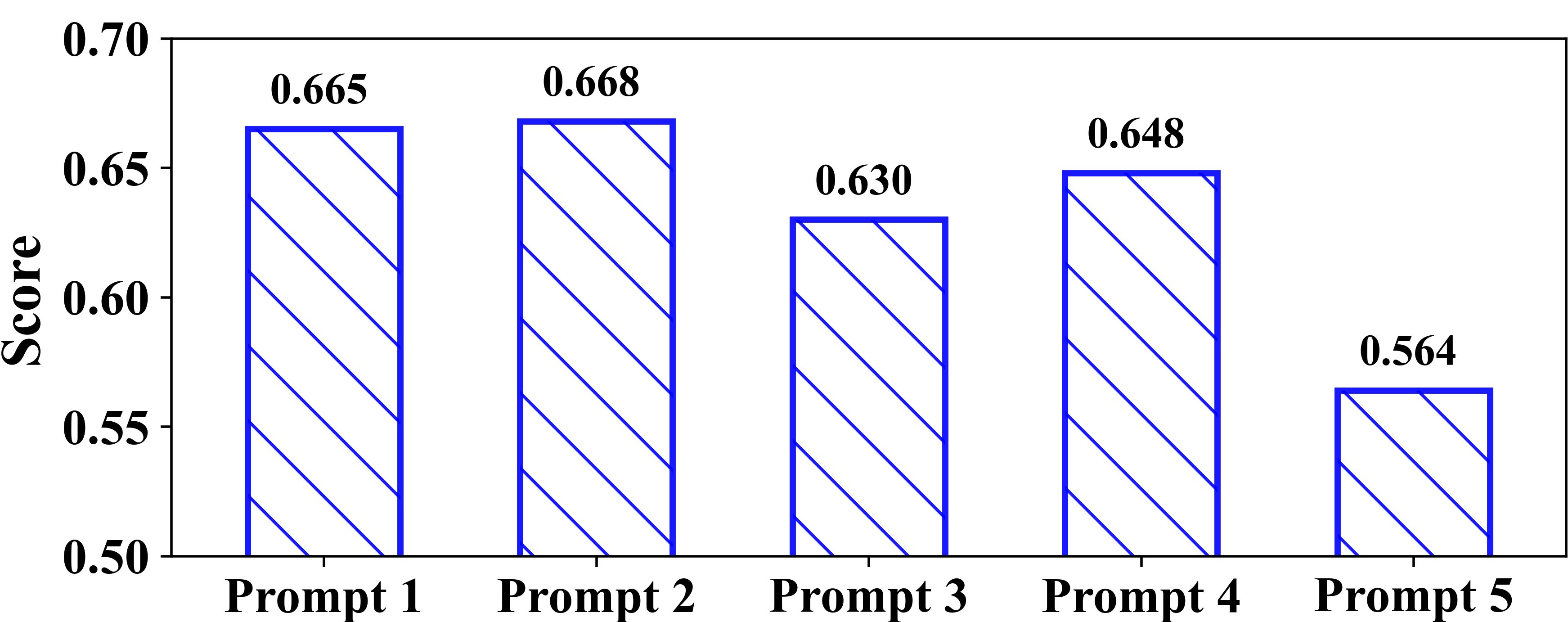}
  \caption{Performance comparison of the five raised prompt candidates in Fig. \ref{fig4} for Log Parsing.}
  \label{fig6}
\end{figure}
Specifically, we evaluated the log parsing performance of the five prompt candidates in Fig. \ref{fig4}, using the first 100 logs of the Android dataset as a test set. The Android dataset has the highest number of log templates (as shown in Table \ref{tab:dataset}), making it suitable for the selection of prompt candidates. The metric score $s(\cdot)$ was set to be the average of RandIndex and F1-score. As shown in Fig.~\ref{fig6}, although ChatGPT considered all five candidate prompts to be suitable for log parsing, a significant fluctuation in performance was observed among the candidates. For example, prompt 2 achieved the best performance, outperforming prompt 5 by 18.4\%. Thus, prompt 2 was chosen to form the complete self-prompt in the main experiment. Note that since Prompt 1 and Prompt 2 achieved similar scores, they both can be regarded as optimal prompts for log parsing in practical applications. A further finding is that prompts that use more accurate language and provide step-by-step instructions tend to produce better outcomes. For example, compared to prompt 5, prompt 2 provides more formal and accurate words (such as ``standardized'' and ``convert'') and clearly outlines the intermediate steps for the task (identifying and replacing variables, then converting to a template). Similar reasoning can be applied to the relative performance of prompts 3 and 4, which outperformed prompt 5 by using more accurate words (such as ``common structure'', ``transform'' and ``variable'').

\subsubsection{CoT Prompt}\label{sec:CoT_prompt} The concept of chain of thought (CoT), a series of intermediate reasoning steps, is introduced by Wei \emph{et al.} \cite{wei2022chain}. The CoT prompt emulates the human thought process when addressing complex problems and can enhance the performance of LLMs in challenging tasks, such as solving mathematical problems. In manual log analysis, practitioners also engage in a series of reasoning steps to reach a conclusion. For instance, without further definitions, the boundary between a normal log and an abnormal log is unclear. Our preliminary experiments revealed that when the simple prompt in Fig. \ref{fig2} is used, ChatGPT often reports anomalies based on hallucinatory content (\emph{e.g.}, 0x00000000, a string of zeros, leads ChatGPT to believe this log recorded an invalid memory access) or reports anomalies based on the belief that a log is abnormally corrupted. Therefore, the intermediate thought process should be regulated in the prompt to suppress the divergence of generated answers.

The CoT prompt guides LLMs in performing log analysis with constrained intermediate steps, both \textit{implicitly} and \textit{explicitly}. (1) \textit{Implicitly}: We require the LLM to generate a reason for each normal/abnormal answer, justifying its decisions. During the pre-training process, the model learned to follow correct and sound logic chains widely existing in cleaned real-world corpora (\emph{e.g.}, wiki pages and books) \cite{brown2020language,ouyang2022training,touvron2023llama}. Our intuition is that, by requesting reasons, the model ``recalls'' the logic chains in training data and implicitly creates a reasonable chain of thoughts, resulting in more logical answers. (2) \textit{Explicitly}: We explicitly define intermediate steps in the prompt. For example, in the task of anomaly detection, we constrain the definition of anomaly to be only ``alerts explicitly expressed in textual content'' to prevent LLMs from ``overthinking'' (\emph{e.g.}, when information is lacking, ChatGPT assumes that the key information is abnormally erased). It is important to note that we include a step to ignore $\langle*\rangle$, the wildcard symbol representing variable parts in a parsed log template, since inputs of this task are log templates.

Assuming $S_a=$ ``a binary choice between abnormal and normal'' as the answer range string, the complete CoT prompt for anomaly detection with format controls is shown as:
\begin{mybox}
$CoT\ prompt$: Classify the given log entries into normal and abnormal categories. Do it with these steps: (a) Mark it normal when values (such as memory address, floating number and register value) in a log are invalid. (b) Mark it normal when lack of information. (c) Never consider $\langle*\rangle$ and missing values as abnormal patterns. (d) Mark it abnormal when and only when the alert is explicitly expressed in textual content (such as keywords like error or interrupt). Concisely explain your reason for each log. $f_x(\texttt{[X]})\ f_z(\texttt{[Z]},S_a)$
\end{mybox}
\subsubsection{In-context Prompt} LLMs are capable of in-context learning, performing a new task through inference alone by conditioning on a few input-label pairs (demonstrations) and making predictions for new inputs \cite{min2022rethinking,brown2020language}. The input-label pairs are directly integrated into the prompt, without requiring an iterative training process (\emph{i.e.}, multiple epochs of parameter updating) and additional resource costs, thereby enabling efficient adaptions in the online scenario. The in-context prompt in LogPrompt adheres to this idea, leveraging a few labelled logs at the beginning of each prompt to create the task context. LLMs are then prompted to provide predictions for logs in the slot \texttt{[X]} based on the given demonstrations in the context. For instance, in the task of anomaly detection, we provide a few randomly sampled normal/abnormal logs and their corresponding categories (1 for abnormal logs and 0 for normal logs) and expect LLMs to implicitly create a task context for anomaly detection and correctly classify input logs. Similarly for log parsing, randomly sampled log-template pairs are used to build the prompt.

Given a small number $m$, $m$ log-label pairs $L_1...L_m$ are randomly sampled. For anomaly detection, with $S_c=$ ``a binary choice between 0 and 1'', the in-context prompt is expressed as:
\begin{mybox}
$In$-$context\ prompt$: Classify the given log entries into 0 and 1 categories based on semantic similarity to the following labelled example logs: (1) Log: $L_1$ Category: 1 (2) Log: $L_2$ Category: 0 \ldots (m-1) Log: $L_{m-1}$ Category: 1 (m) Log: $L_{m}$ Category: 0. $f_x(\texttt{[X]})\ f_z(\texttt{[Z]},S_c)$
\end{mybox}
For log parsing, the prompt can be built in a similar manner with the corresponding template attached behind each example log and $S_c$ replaced with the one in Section~\ref{sec:self_prompt}.

\section{Experiments and Evaluations}\label{sec:experiment}
In this section, we first discuss the experimental setup. We then propose four research questions and discuss the results.
\subsection{Experimental Setup}
\subsubsection{Datasets}
\begin{table}[htbp]
\caption{The Statistics of Log Datasets in the Experiments.}
\centering
\resizebox{\linewidth}{!} {%
\begin{tabular}{lrccp{0.06\linewidth}<{\centering}p{0.06\linewidth}<{\centering}}
\toprule
\textbf{Dataset}  & \textbf{\#Messages} & \textbf{\#Templates (2k)} & \textbf{\#Anomalies} & \textbf{L.P.$^{\mathrm{a}}$} & \textbf{A.D.$^{\mathrm{a}}$}\\
\midrule
HDFS   &  \num{82293702}  & \num{14} & - & \Checkmark &  - \\
Hadoop  &  \num{394308}  & \num{114} & - & \Checkmark &  - \\
Zookeeper    &   \num{74380}  & \num{50} & - &\Checkmark&  - \\
BGL    &    \num{4713493} & \num{120} & \num{348460} & \Checkmark &  \Checkmark\\
HPC   &    \num{433489} & \num{46} & - & \Checkmark &  - \\
Linux   &    \num{25567} & \num{118} & -  & \Checkmark & - \\
Proxifier  & \num{21329} & \num{8} & - & \Checkmark &  -\\
Android  & \num{30348042} & \num{166} & - & \Checkmark & - \\
Spirit  &  \num{7958733} & - & \num{768142} & - & \Checkmark \\
\bottomrule
\multicolumn{6}{l}{$^{\mathrm{a}}$ \textbf{L.P.} denotes the task of log parsing, \textbf{A.D.} denotes anomaly detection.}\\
\multicolumn{6}{l}{~~\,\Checkmark means the dataset is used as a task evaluation set with human labels. }\\
\end{tabular}
}
\label{tab:dataset}
\end{table}

As shown in Table \ref{tab:dataset}, the effectiveness of LogPrompt is evaluated on large-scale real-world log datasets from nine different domains, including supercomputers, distributed systems, operating systems, mobile systems, and server applications \cite{he2020loghub}. Two of the datasets were annotated by domain experts to identify anomalous events for the purpose of anomaly detection \cite{oliner2007supercomputers}. The number of labelled anomalies can be found in Table \ref{tab:dataset}. To evaluate the log parsing performance, eight of the datasets have log templates manually extracted from a subset of 2000 log messages in each domain \cite{zhu2019tools}. All log data were timestamped, enabling the datasets to be split into train/test sets chronologically.

\subsubsection{Implementation Details}\label{sec:detail}
In our primary experiments, the underlying LLM is ChatGPT and is accessed via APIs provided by OpenAI (model version: \textit{gpt-3.5-turbo-0301}). In Section \ref{sec:more_llms}, an open-source LLM with fewer parameters is deployed using 1 NVIDIA A100 GPU. The initial temperature coefficient of ChatGPT is set to 0.5, maintaining a balance by increasing the model's reasoning capabilities through diverse token exploration while limiting detrimental randomness~\cite{xu2022systematic}. If the response format is invalid, the query is resubmitted with an increased temperature coefficient of 0.4 until the response format is correct. The format failure rate is less than 1\%, which is consistent with \cite{kocmi2023large}. The train/test datasets are split chronologically to simulate online scenarios.
\subsection{Research Questions and Results}
\subsubsection{RQ1: How effective is LogPrompt compared with existing methods in the online scenario, when being faced with scarce training data?}\label{sec:RQ1}

\paragraph{Motivation.} As shown in Figure \ref{fig3}, the performance of existing log analysis methods is often reduced in the online scenario, where available in-domain training logs are often limited. The aim of this Research Question (RQ) is to investigate whether LogPrompt can address this challenge. In this comparison experiment, LogPrompt stands out for its unique approach of not requiring in-domain training, while existing approaches are trained on a small portion (less than 10\%) of in-domain logs to simulate the online scenario. The amount of training data we provided for existing approaches aligns with previous studies on online log analysis \cite{meng2020logparse,tao2022logstamp,biglog}. It should be noted that, as will be discussed in Section~\ref{sec:threat}, the performance of some approaches may be underestimated in this setting. 

\paragraph{Approach.} To evaluate the performance of LogPrompt for log parsing, we compare it against 10 existing methods, namely LogPPT \cite{le2023log}, Logstamp \cite{tao2022logstamp}, LogParse \cite{meng2020logparse}, LogSig \cite{tang2011logsig}, Spell~\cite{du2016spell}, IPLoM \cite{makanju2009clustering}, Drain \cite{he2017drain}, FT-tree \cite{zhang2017syslog}, MoLFI \cite{messaoudi2018search}, and LKE \cite{fu2009execution}. The datasets used in the evaluation are described in Table \ref{tab:dataset}. For each dataset, most baseline methods are trained on the first 10\% logs and evaluated on the remaining 90\% logs, while LogPrompt is directly tested on the remaining 90\% logs. Note that LogPPT \cite{le2023log} is a recent log parsing approach specialized for few-shot scenarios. Thus, we set the ratio of its training data to 0.05\%. 

In our evaluation of log parsing, we adopt the F1-score as the metric. While RandIndex~\cite{rand1971objective} is commonly used, it only assesses the accuracy of log clustering and does not consider the correctness of variables in the extracted templates. Recent studies~\cite{da-parser,biglog} have highlighted that RandIndex is not sensitive to parsing errors and may assign a high score even to random clustering. In contrast, the F1-score presents a more challenging metric as it requires precise identification of variable parts in logs~\cite{huo2023semparser,li2023did}, which is particularly crucial for parsing unseen logs in online scenarios. To calculate the F1-score, we tokenize the predicted log template into a list of tokens. We then define the terms $TP$, $TN$, $FP$, and $FN$ as follows: (1) $TP$: the token is a variable and is correctly predicted as such, (2) $TN$: the token is not a variable and is predicted not to be a variable, (3) $FP$: the token is not a variable but is mistakenly predicted as a variable, and (4) $FN$: the token is a variable but is predicted not to be a variable. The F1-score is computed as the harmonic average of Recall ($Recall = \frac{TP}{TP+FN}$) and Precision ($Precision = \frac{TP}{TP+FP}$).

For the task of anomaly detection, we compare LogPrompt with three representative existing methods: DeepLog~\cite{du2017deeplog}, LogAnomaly \cite{meng2019loganomaly}, and LogRobust~\cite{zhang2019robust}. Following the paradigm of anomaly detection raised in~\cite{le2022log}, we first parse each input log into a log template and form log sessions from templates, using fixed-window grouping with a length of 100 templates. To ensure a fair comparison, for all methods, the templates are extracted using Drain \cite{he2017drain}, a fast log parsing algorithm. For baselines, the first 4000 logs in each dataset are used for training. Both LogPrompt and the trained baselines are then tested on the remaining logs. For logPrompt, the LLM first predicts a normal/abnormal tag for each log template in a session and the session is considered abnormal if any abnormal tag exists. In this experiment, since existing methods output only session-level anomalies, we use the F1-score for session anomaly as the evaluation metric (See template-level performance in Table~\ref{tab:CoT_prompt}), where $TP$ represents the successful identification of an abnormal session (similarly for $TN$, $FP$, and $FN$). We also report the Recall and Precision values. For the forecast-based methods (DeepLog and LogAnomaly), we report the top-4 F1-score.

\begin{table}[tbp]
\caption{F1-score of Log Parsing in the Online Scenario.}
\centering
\resizebox{\linewidth}{!} {%
\begin{tabular}{l@{\hskip 0.05in}c@{\hskip 0.05in}c@{\hskip 0.05in}c@{\hskip 0.05in}c@{\hskip 0.05in}c@{\hskip 0.05in}c@{\hskip 0.05in}c@{\hskip 0.05in}c@{\hskip 0.05in}||c}
\toprule
\textbf{Methods$^{\mathrm{a}}$}                  & \textbf{HDFS}   & \textbf{Hd$^{\mathrm{b}}$} & \textbf{Zk} & \textbf{BGL}    & \textbf{HPC}    & \textbf{Linux}  & \textbf{Px} & \textbf{Andro} &\textbf{Avg.}   \\ \midrule
\noalign{\vskip 2pt}
IPLoM \cite{makanju2009clustering}                   & 0.389          & 0.068          & 0.225             & 0.391          & 0.002          & 0.225          & 0.500     &    0.419    & 0.277          \\
LKE \cite{fu2009execution}                     & 0.424          & 0.198          & 0.225             & 0.379          & 0.381          & 0.388          & 0.309      & 0.000       & 0.288          \\
LogSig \cite{tang2011logsig}                  & 0.344          & 0.050          & 0.225             & 0.333          & 0.002          & 0.146          & 0.339      &   0.116    & 0.194          \\
FT-tree \cite{zhang2017syslog}                 & 0.385          & 0.046          & 0.186             & 0.497          & 0.002          & 0.211          & 0.420       &    0.581  & 0.291          \\
Spell \cite{du2016spell}                   & 0.000          & 0.058          & 0.045             & 0.536          & 0.000          & 0.091          & 0.000      &    0.245    & 0.122          \\
Drain \cite{he2017drain}                   & 0.389          & 0.068          & 0.225             & 0.397          & 0.002          & 0.225          & 0.500     &    0.413    & 0.277          \\
MoLFI \cite{messaoudi2018search}                   & 0.000          & 0.095          & 0.000             & 0.333          & 0.000          & 0.026          & 0.000     &   0.208    & 0.083          \\
LogParse \cite{meng2020logparse}                & 0.632          & 0.502          & 0.348             & 0.665          & 0.330          & 0.588          & 0.334      &    0.233   & 0.454          \\
LogStamp \cite{tao2022logstamp}                & 0.523          & 0.594          & 0.275             & 0.818          & 0.434          & 0.658          & 0.438      &    \textbf{0.899}   & 0.580          \\
LogPPT \cite{le2023log}                   & 0.838 & 0.526 & 0.795    & \textbf{0.982} & 0.287 & 0.423 & 0.638    & 0.313 & 0.600 \\
\hdashline
\noalign{\vskip 3pt}
\textbf{LogPrompt} &  \textbf{0.863} & \textbf{0.763} &  \textbf{0.889} & 0.865 &  \textbf{0.759}  & \textbf{0.766} & \textbf{0.653} & 0.819 & \textbf{0.797}\\ \bottomrule
\multicolumn{10}{l}{$^{\mathrm{a}}$ LogPrompt uses \textbf{no training data}, unlike others trained with up to 10\% logs.}\\
\multicolumn{10}{l}{$^{\mathrm{b}}$ \textbf{Hd}, \textbf{Zk}, \textbf{Px} and \textbf{Andro} denotes Hadoop, Zookeeper, Proxifier and Android.}\\
\end{tabular}
}
\label{tab:logParsing_zeroShot}
\end{table}

\paragraph{Result of Log Parsing.} The results of the log parsing task are presented in Table \ref{tab:logParsing_zeroShot}. Utilizing the self-prompt strategy in Section \ref{sec:self_prompt}, LogPrompt achieved the best F1-score on six of the eight datasets. Moreover, despite not being trained on in-domain logs, LogPrompt outperformed LogPPT by 32.8\% and LogStamp by 37.4\% in average, which require resources for in-domain training. The advantage of LogPrompt indicates its strong ability in accurately recognizing variables from unseen logs. It is worth noting that, according to recent studies \cite{huo2023semparser,li2023did}, accurately identifying fine-level variables in logs is of greater significance in the analysis of breakdowns and failures in many industrial scenarios, as opposed to solely relying on coarse-level clustering of logs. The effectiveness of LogPrompt in recognizing fine-level variables in logs makes it a suitable choice for parsing unseen logs in online scenarios.

\begin{table}[htbp]
\caption{Log Anomaly Detection in the Online Scenario.}
\centering
\resizebox{\linewidth}{!} {%
\setlength{\tabcolsep}{0pt}
\begin{tabular}{l@{\hskip 0.1in}c@{\hskip 0.1in}c@{\hskip 0.1in}c@{\hskip 0.1in}c@{\hskip 0.1in}c@{\hskip 0.1in}c@{\hskip 0.1in}c}
\toprule
\multirow{3}{*}{\textbf{Methods}}  & \multirow{3}{*}{\textbf{\#Train}$^{\mathrm{a}}$} & \multicolumn{3}{c}{\textbf{BGL}$^{\mathrm{c}}$}                        & \multicolumn{3}{c}{\textbf{Spirit}}         \\ \cmidrule(l{-0.3em}r{1.3em}){3-5} \cmidrule(l{-0.3em}r){6-8} 
                               &  & \textbf{P} & \textbf{R} & \textbf{F}  & \textbf{P} & \textbf{R} & \textbf{F}         \\ \midrule
DeepLog \cite{du2017deeplog}               &               & 0.156             & \textbf{0.939}          & 0.268          & 0.249             & 0.289          & 0.267                  \\
LogAnomaly \cite{meng2019loganomaly}        &            & 0.016             & 0.056          & 0.025          & 0.231             & 0.141          & 0.175                  \\
LogRobust \cite{zhang2019robust}            &      \multirow{-3}{*}{{4000}}        & 0.095             & 0.425          & 0.156      & 0.109             & 0.135          & 0.120               \\
\hdashline
\noalign{\vskip 2pt}
\textbf{LogPrompt}$^{\mathrm{b}}$           &      \textbf{0}           & \textbf{0.249}             & 0.834          & \textbf{0.384}        & \textbf{0.290}             & \textbf{0.999}          & \textbf{0.450}  \\ \bottomrule
\multicolumn{8}{l}{$^{\mathrm{a}}$ \textbf{\#Train} denotes the number of logs utilized for training.}\\
\multicolumn{8}{l}{~~\,\textbf{P} denotes Precision. \textbf{R} denotes Recall. \textbf{F1} denotes F1-score.}\\
\multicolumn{8}{l}{$^{\mathrm{b}}$ LogPrompt here is constructed via the strategy of CoT-prompt.}\\
\multicolumn{8}{l}{$^{\mathrm{c}}$ We automatically extracted 1766 templates from BGL; 1297 from Spirit.}\\
\multicolumn{8}{l}{~~\,The reported F1-score is session level. See template level in Table \ref{tab:CoT_prompt}.}\\
\end{tabular}
}
\label{tab:anomaly_zeroShot}
\end{table}

\paragraph{Result of Anomaly Detection.} The results of the anomaly detection task are presented in Table \ref{tab:anomaly_zeroShot}. Despite the existing methods being trained on thousands of logs, LogPrompt still achieved strong performances in both datasets without utilizing any in-domain training data, with an average improvement of 55.9\% in terms of F1-score. This indicates the effectiveness of LogPrompt in addressing the performance reduction in anomaly detection when training data is scarce. However, even with LogPrompt, the absolute performance of anomaly detection in the online scenario remains low, highlighting the challenging nature of performing anomaly detection with extremely limited historical in-domain logs.

\subsubsection{RQ2: Does LogPrompt's interpretability yield helpful and comprehensible content for practitioners during real-world log analysis?}

\begin{table*}[htbp]
\caption{Human Scoring Criteria for Evaluating the Interpretability of Large Language Models in Log Analysis. Score Ranking: 1 - Not Interpretable, 2 - Low Interpretable, 3 - Moderately Interpretable, 4 - Highly interpretable, 5 - Very Highly interpretable.}
\centering
\resizebox{\linewidth}{!} {%
\begin{tabular}{p{0.02\linewidth}p{0.36\linewidth}p{0.33\linewidth}p{0.32\linewidth}}
\toprule
\multirow{2}{*}{\textbf{Scores}}                                     & \multicolumn{2}{c}{\textbf{Usefulness}}                     & \multicolumn{1}{c}{\multirow{2}{*}{\textbf{Readability}}}  \\\cmidrule(l{1.5em}r{1.5em}){2-3}
 & \makecell[c]{\textbf{Log Parsing}} &  \makecell[c]{\textbf{Anomaly Detection}} & \\
\midrule
1     & No variable is extracted, or explanations of those variables are irrelevant.          &   No justification on the anomaly beyond a simple prediction label.  & The text contains numerous unintelligible elements or grammatical mistakes.                      \\
\midrule
2     & Explanations are meaningless or logically wrong, hindering engineers from interpreting the logs.          &  The Justification for the prediction is irrelevant, or logically inconsistent with the facts.   & Most of the generated text is readable, but it may have grammar errors or unclear phrases.  \\
\midrule
3     & Variables with appropriate classes are partially extracted, and explanations are somewhat relevant.          & The justification well supports the predictions, but may lack clarity and details. & The text has few grammar errors, although some terms may need refinement. \\
\midrule
4    &   Variables with appropriate classes are mostly extracted, and explanations are specific and relevant. Engineers can understand the logs with less effort.      & Specific, accurate, and relevant justification is presented, which positively assist engineers in eliminating false alarms and further analysis. &The text is clear, grammatically correct, with only a minimal number of technical terms possibly needing refinement. \\
\midrule
5    & Variables are fully and correctly extracted, and explanations are detailed, specific and relevant, enabling easy and precise understanding of logs.         & Detailed, relevant and clear justification that significantly assists engineers in ruling out false alarms and locating the root cause.  & The text is clear, detailed, grammatically perfect, and professional for software engineering. \\
\bottomrule
\end{tabular}
}
\label{tab:Scoring_Criteria}
\end{table*}

\paragraph{Motivation.} As discussed in Section \ref{sec:motivation}, the insufficient interpretability of existing approaches is a major pain point for practitioners during log analysis. Although justifications and explanations, as shown in Fig. \ref{LogAnalysis}(b), can be generated by LogPrompt, it is still hard to determine the quality of such content in industrial scenarios. To assess the impact of LogPrompt on enhancing the interpretability of log analysis, we established a set of criteria for evaluating interpretability, presented in Table \ref{tab:Scoring_Criteria}. We cooperated with a top-tier ICT \& software company and recruited six of their experienced practitioners (with full-time commitment to this project) to evaluate LogPrompt's interpretability in terms of usefulness and readability. The recruited experts possess over 10 years of experience of O\&M in various domains such as distributed systems, mobile operating systems, and software services.

\paragraph{Approach.} A total of 200 logs were randomly sampled for the human evaluation, accompanied by LogPrompt's actual outputs. Specifically, 100 logs were related to log parsing, evenly distributed across the eight domains that have human-labelled variables and templates in Table \ref{tab:dataset}. While the remaining 100 were related to anomaly detection, evenly distributed across BGL and Spirit, the two domains with human labels. As the correctness of LogPrompt was already evaluated in Section \ref{sec:RQ1}, incorrectly predicted logs (\emph{i.e.}, FPs and FNs) were not included in this evaluation. To ensure a comprehensive evaluation, an equal number of normal and abnormal samples were included for anomaly detection, and each selected log for log parsing was required to contain at least one variable. Fig. \ref{LogAnalysis}(b) presents two example formats of logs with LogPrompt's outputs, which were rated by the reviewers. For anomaly detection, we utilized the CoT prompt (implicit part only) to prompt ChatGPT to generate task outputs along with justifications. For log parsing, we constrained the variables to the eight general categories defined by Li \emph{et al.} \cite{li2023did} and prompted ChatGPT to classify and generate explanations for each variable. Following Ahmed \emph{et al.}~\cite{ahmed2023recommending}, the evaluation dimensions focused on the following two aspects:
\begin{itemize}
\item \textbf{Usefulness.} The reviewers were asked to rate the extent to which the provided interpretation of the analysis decision is detailed, specific, relevant, logically sound and helpful in actual log analysis.
\item \textbf{Readability.} The reviewers were asked to rate the ease with which a reader could understand the provided text. A text was considered readable if it was grammatically correct, meaningful, and professional.
\end{itemize}

For each evaluation dimension, a Likert scale ranging from one to five is utilized to score the interpretability level, with Table \ref{tab:Scoring_Criteria} providing detailed definitions for each level. Every reviewer independently rated all the 200 samples. For each log, the reviewers were asked to provide a usefulness score and a readability score based on their understanding of the scoring criteria. To mitigate bias, the reviewers were unaware of the source of the generated explanations during evaluations, and there was no overlap between co-authors and reviewers. We reported two metrics: (1) The average score assigned by each reviewer across all samples. (2) The High Interpretability Percentage (HIP), which represents the proportion of samples that received a score higher than four from a reviewer.

\begin{table}[htbp]
\caption{Interpretability of LogPrompt Rated by Experts.}
\centering
\resizebox{\linewidth}{!} {%
\setlength{\tabcolsep}{0pt}
\begin{tabular}{l@{\hskip 0.1in}c@{\hskip 0.1in}c@{\hskip 0.1in}c@{\hskip 0.1in}c@{\hskip 0.1in}c@{\hskip 0.1in}c@{\hskip 0.1in}c@{\hskip 0.1in}c}
\toprule
\multirow{3}{*}{\vspace{-1.5em}\textbf{Raters}}  & \multicolumn{4}{c}{\hspace{-1em}\textbf{Log Parsing}}                        & \multicolumn{4}{c}{\textbf{Anomaly Detection}}         \\ \cmidrule(l{-0.3em}r{1.5em}){2-5} \cmidrule(l{-0.3em}r{-0.01em}){6-9} 
   & \multicolumn{2}{l}{\hspace{0.5em}\textbf{Usefulness}}            & \multicolumn{2}{l}{\hspace{0.3em}\textbf{Readability}} & \multicolumn{2}{l}{\hspace{0.5em}\textbf{Usefulness}}            & \multicolumn{2}{l}{\hspace{0.3em}\textbf{Readability}}        \\ \cmidrule(l{-0.1em}r{1.6em}){2-3} \cmidrule(l{-0.1em}r{1.6em}){4-5} \cmidrule(l{-0.1em}r{1.6em}){6-7} \cmidrule(l{-0.1em}r){8-9}
                               & \textbf{Mean}$^{\mathrm{a}}$ & \textbf{HIP}$^{\mathrm{b}}$ & \textbf{Mean} & \textbf{HIP}   & \textbf{Mean} & \textbf{HIP} & \textbf{Mean} & \hspace{-0.2em}\textbf{HIP}         \\ \midrule
R1               &       4.19        & 76.00\%      &   4.70    & 94.00\%          & 4.27     &  73.00\%   & 4.50             & 84.00\%                \\
R2        &      4.36      & 81.00\%             & 4.78          & 95.00\%         & 4.42     &    79.00\%    & 4.63          & 90.00\%                 \\
R3        &      4.13       & 91.00\%             & 4.18          & 95.00\%     &   4.12  & 86.00\%             & 4.35         & 94.00\%               \\
R4        &      4.20       & 73.00\%             & 4.74          & 98.00\%     &   4.40  & 85.00\%             & 4.71         & 95.00\%               \\
R5        &      4.06       & 91.00\%             & 4.41          & 100.00\%     &   4.12  & 94.00\%             & 4.24         & 98.00\%               \\
R6        &      4.46       & 86.00\%             & 4.77          & 97.00\%     &   4.48  & 90.00\%             & 4.78         & 99.00\%               \\
\hdashline
\noalign{\vskip 3pt}
\textbf{Avg.}  &      4.23       & 83.00\%             & 4.60          & 96.50\%     &   4.30  & 84.50\%             & 4.54         & 93.33\% \\
\bottomrule
\multicolumn{9}{l}{$^{\mathrm{a}}$ \textbf{Mean} is the average score of the ratings by a reviewer for all samples.}\\
\multicolumn{9}{l}{$^{\mathrm{b}}$ \textbf{HIP} is the percentage of samples received a rating higher than four.}\\
\end{tabular}
}
\label{tab:interpretability_result}
\end{table}

\paragraph{Result.} The results of the human evaluation are presented in Table \ref{tab:interpretability_result}. For each reviewer, the average scores on the samples for both tasks in terms of usefulness and readability consistently exceeded four. For the HIP, there were some fluctuations ranging from 70\% to 90\%, which can be attributed to the subjective nature of the criteria. However, in every evaluation dimension, the average HIP was consistently above 80\%, indicating that most generated content was considered helpful and readable by experienced log analysts.

\paragraph{Feedbacks from Practitioners.} Alongside the ratings, we obtained valuable comments and feedback from the reviewers. One practitioner states that: ``I appreciate the ability of LogPrompt in supporting the interpretation of logs from various domains. As our system continuously incorporating third-party services, sometimes I have to read through the manuals to decipher logs from unfamiliar domains. The explanations generated by LogPrompt can provide a swift grasp of the logs before I found the official definitions.'' 

Another comments said: ``LogPrompt can definitely help in the compose of analysis reports, especially shortly after system crashes, where I often need to quickly interpret the events to non-technical colleagues in meetings.'' 

Another reviewer commented: ``In the realm of software O\&M, false alarms are an inescapable reality, with false positives imposing substantial time costs on engineers and false negatives causing severe ramifications. Accompanying explanations with automatic analysis outcomes enables engineers to more promptly ascertain the credibility of the purported anomaly, thereby reducing the time spent on subsequent actions.''

\begin{table}[htbp]
\caption{Analysis on low-rated samples in human evaluation }
\centering
\resizebox{\linewidth}{!} {%
\begin{tabular}{p{0.17\linewidth}p{0.06\linewidth}p{0.28\linewidth}p{0.34\linewidth}}
\toprule
\multicolumn{1}{l}{\multirow{2}[3]{*}{\textbf{Cause}}} & \multicolumn{1}{c}{\multirow{2}[3]{*}{\textbf{Ratio}}} & \multicolumn{2}{c}{\textbf{Example}}\\
\cmidrule(lr){3-4}
 &  & \makecell[c]{\textbf{Raw log}} & \makecell[c]{\textbf{Interpretation}}\\
\midrule
\multirow{3}{*}{\begin{tabular}[c]{@{}l@{}}Domain \\Knowledge \end{tabular}} & \multirow{3}{*}{45.95\%} & CE sym 33, at 0x1ff2fc60, mask 0x04. &  Sym 33 is a Object Name that refers to an object named ``sym 33''.\\
\midrule
\multirow{3}{*}{\begin{tabular}[c]{@{}l@{}}Semantic \\insufficiency \end{tabular}} & \multirow{3}{*}{40.54\%} & fpr8=0x3883a497 bfe016b5 8da3311e 3f06d5a7. &  \multirow{3}{*}{\begin{tabular}[c]{@{}l@{}}Normal - Register\\values. \end{tabular}}\\
\midrule
Other & 13.51\% & \makecell[c]{-} & \makecell[c]{-}\\
\bottomrule
\end{tabular}
}
\label{tab:badCase}
\end{table}

\paragraph{Bad Case Analysis.} We investigate the samples that received ratings averagely lower than four by practitioners. As shown in Table \ref{tab:badCase}, a major factor is the LLM's lack of domain knowledge, which leads to overly general interpretations of some domain-specific terms. For example, it may refer to specific parameters as simply an object. Another cause is the lack of semantic contents in certain input logs, which can be attributed to their brevity or richness of non-NLP patterns (\emph{i.e.}, digits, codes and addresses). This insufficiency makes it challenging for the model to generate sound explanations.                                                                                                             

The practitioners' feedback on LogPrompt indicated its potential for real-world applications and provided valuable insights for optimizing the model, including the need to improve the model's domain-specific knowledge and its understanding of non-NLP patterns in logs through domain adaptation technologies.

\subsubsection{RQ3: How do varying prompt strategies within LogPrompt impact the performance of LLMs in log analysis tasks?}\label{sec:ablationStudy}

\paragraph{Motivation.} Given the prompt strategies discussed in Section \ref{sec:prompt_strategy}, this RQ aims to investigate the influence of these strategies on the performance of LLMs in log analysis tasks. Specifically, this RQ seeks to explore the impact on performance by examining various settings and modules in LogPrompt, which will help guide the selection of optimal settings for LogPrompt in practical applications and enhance understanding of the mechanism of LogPrompt.

\begin{table}[htbp]
\caption{Ablation Study of the CoT Prompt Strategy in Anomaly Detection.}
\centering
\begin{tabular}{ccccc}
\toprule
\multirow{3}{*}{\textbf{Methods}$^{\mathrm{b}}$}  & \multicolumn{2}{c}{\textbf{BGL}}                        & \multicolumn{2}{c}{\textbf{Spirit}}         \\ \cmidrule(lr){2-3} \cmidrule(lr){4-5} 
                               & \textbf{S-F1}$^{\mathrm{a}}$ & \textbf{T-F1}   & \textbf{S-F1} & \textbf{T-F1}         \\ \midrule
ChatGPT (simple prompt) & 0.189 & 0.115 & 0.445 & 0.086  \\
~~\,+Implicit CoT & 0.307 & 0.183 & 0.443 & 0.096   \\
~~\,+Explicit CoT & \textbf{0.384} & \textbf{0.321} & \textbf{0.450} & \textbf{0.116}  \\
\bottomrule
\multicolumn{5}{l}{$^{\mathrm{a}}$ \textbf{S-F1} denotes the session-level F1-score.}\\
\multicolumn{5}{l}{~~\,\textbf{T-F1} denotes the template-level F1-score.}\\
\multicolumn{5}{l}{$^{\mathrm{b}}$ The implicit CoT requires ChatGPT to report justifications.}\\
\multicolumn{5}{l}{~~\,The explicit CoT further adds intermediate steps in prompts.}\\
\end{tabular}
\label{tab:CoT_prompt}
\end{table}

\paragraph{Impact of Modules in the CoT Prompt.} As detailed in Section \ref{sec:CoT_prompt}, the CoT prompt consists of two modules: implicit CoT and explicit CoT. Starting from the simple prompt in Fig. \ref{fig2}, we successively added each module to evaluate its impact on the performance of ChatGPT in the task of anomaly detection. In addition to the session-level F1-score reported in Table \ref{tab:anomaly_zeroShot}, we also report the template-level F1-score, which reflects the performance in predicting the anomaly tag for each log template. The results of our evaluation are presented in Table \ref{tab:CoT_prompt}, which demonstrate that incorporating chains of thought in the prompt continuously improves the performance of ChatGPT in anomaly detection (\emph{e.g.}, for the BGL dataset, the session-level F1-score improved by 0.195 and the template-level F1-score improved by 0.206). This supports the idea that a more precise and step-by-step prompt leads to better performance from LLMs.

An interesting observation from our evaluation is that even when only the implicit part of the CoT prompt was utilized, the performance of ChatGPT was still significantly improved (\emph{e.g.}, for the BGL dataset, the session-level F1-score improved by 62.4\% and the template-level F1-score improved by 59.1\%). The only difference between the simple prompt and the prompt loaded with the implicit CoT was that the LLM was asked to generate justifications for its analysis decisions. This improvement in performance can be attributed to the fact that the LLM implicitly generates reasonable intermediate steps for the task, leading to a more logical result when being prompted to justify its decisions. Given that language models are trained to fit sound logical chains from various real-world corpora, generating extra explanations in log analysis results reduces the likelihood of the language model inferring incorrect logical chains. This further emphasizes the importance of interpretability in log analysis.

\begin{figure}[htbp]
    \centering
  \includegraphics[width=0.9\linewidth]{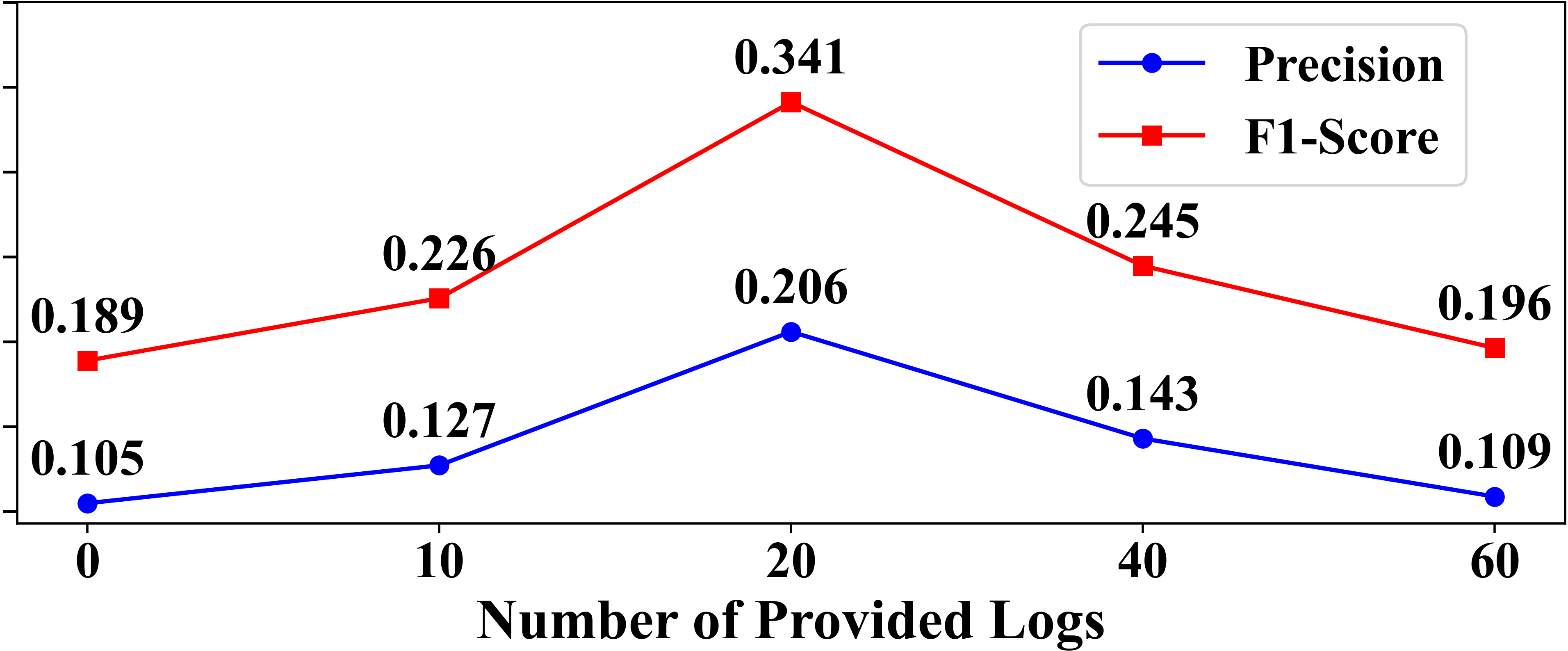}
  \caption{Variation in F1-score and Precision of the in-context prompt with an increase in the number of provided logs.}
  \label{fig5}
\end{figure}

\paragraph{Impact of Number of Labelled Logs in the In-context Prompt.} Fig. \ref{fig5} presents the performance of the in-context prompt in the anomaly detection task on the BGL dataset when we vary the number of labelled logs provided in the prompt. The Recall is mostly steady across the number of logs (consistently achieving 0.99+ with the only exception of 0.847 observed when the number of logs is 40), and thereby is not presented in the figure. For each number of logs, the number of positive samples is equal to the number of negative samples. The results demonstrate that even with a few labelled logs provided in the prompt, the performance of ChatGPT is significantly improved compared to the simple prompt (\emph{i.e.}, where no labelled logs are provided), indicating the effectiveness of the in-context prompt. Once given a few log examples that are divided into positive and negative clusters, even without explicitly specifying the context of anomaly, the LLM can classify new input logs into their corresponding clusters, by implicitly creating a task context for anomaly detection. 

As seen from Fig. \ref{fig5}, both the F1-score and precision reach their peak when 20 labelled logs are provided in the prompt, and then decrease with the further increase of labelled logs. This phenomenon can be attributed to the idea that, instead of learning the input-label correspondences from large-scale training data like typical supervised learning, LLMs model other data components (\emph{i.e.}, example inputs, example labels, and the example format) from the examples in the in-context prompt, which can be efficiently learned from only a few examples \cite{min2022rethinking}. Additionally, an overly long context prefixed to the prompt may cause LLMs to pay less attention to the new input logs, thereby deteriorating the task performance.

\subsubsection{RQ4: Is LogPrompt Compatible With Open-source and Smaller-scale LLMs?}\label{sec:more_llms}

\paragraph{Motivation.} Although LogPrompt exhibits promising qualities, it is important to address two key concerns. Firstly, the deployment of large, proprietary, and API-dependent LLMs, such as ChatGPT, in local environments can be a challenging task. Secondly, the reproducibility of LogPrompt may be compromised if the API services of LLMs become unavailable. Therefore, it is crucial for LogPrompt to have compatibility with alternative open-source, privacy-protected, and smaller-scale LLMs. In this RQ, we explore the compatibility of LogPrompt with a smaller-sized and open-source LLM.

\paragraph{Approach.} In addition to ChatGPT, the proprietary model with 175B parameters developed by OpenAI, we further evaluate the performance of LogPrompt under \textit{Vicuna} \cite{vicuna2023}, an open-source LLM with only 13B parameters. \textit{Vicuna} \cite{vicuna2023} can be easily deployed on MacBook and mobile platforms, requiring minimally only CPU and less than 7GB of memory\footnote{An approach of efficient deployment: https://github.com/ggerganov/llama.cpp}. We investigate the performance of log parsing on the eight datasets in Table \ref{tab:logParsing_zeroShot} without in-domain training, which is the same setting as Section \ref{sec:RQ1}.

\begin{table}[htbp]
\caption{Performances of LogPrompt in Online Log Parsing with Varying Underlying LLMs.}
\centering
\resizebox{\linewidth}{!} {%
\setlength{\tabcolsep}{0pt}
\begin{tabular}{lc@{\hskip 0.05in}c@{\hskip 0.05in}c@{\hskip 0.05in}c@{\hskip 0.05in}c@{\hskip 0.05in}c@{\hskip 0.05in}c@{\hskip 0.05in}c@{\hskip 0.01in}c}
\toprule
\textbf{LLMs}  & \textbf{Size}        & \textbf{HDFS}   & \textbf{Hd$^{\mathrm{a}}$} & \textbf{Zk} & \textbf{BGL}    & \textbf{HPC}    & \textbf{Linux}  & \textbf{Px} & \textbf{Andro}        \\ 
\midrule
\multicolumn{10}{l}{\textbf{GPT Model}}  \\
\textit{ChatGPT}     &       \textbf{175B}         &  0.863 & 0.763 &  0.889 & 0.865 &  0.759  & 0.766 & 0.653 & 0.819           \\
\hdashline
\noalign{\vskip 3pt}
 \multicolumn{10}{l}{\textbf{Vicuna$^{\mathrm{a}}$} (with)} \\
\textit{Simple Prompt}   &     &      0.244       & 0.018             & 0.273          & 0.048     &   0.022  & 0.109         & 0.014         & 0.041               \\
\textit{LogPrompt}     &    \multirow{-2}{*}{{\textbf{13B}}} &      0.779       & 0.468             &     0.364      & 0.266     &   0.243  & 0.715        & 0.614         & 0.248               \\
\bottomrule
\multicolumn{10}{l}{$^{\mathrm{a}}$ \textbf{Hd}, \textbf{Zk}, \textbf{Px}, \textbf{Andro} denotes Hadoop, Zookeeper, Proxifier, Android.}\\
\multicolumn{10}{l}{$^{\mathrm{b}}$ Vicuna with LogPrompt uses the in-context prompt strategy with $m$=3. }\\

\end{tabular}
}
\label{tab:varying_LLM}
\end{table}

\paragraph{Result.} The results in Table \ref{tab:varying_LLM} demonstrate that, although Vicuna has only 13B parameters, when equipped with LogPrompt, it achieves a comparable performance of log parsing with the 175B GPT model in the datasets of HDFS, Linux and Proxifier. Additionally, when the prompt strategy transitions from a simple prompt to LogPrompt, Vicuna exhibits significant improvements on performance, with an average increase of 380.7\% in F1-score. The effectiveness of LogPrompt with the significantly smaller-scale LLM highlights its reproducibility and compatibility. Furthermore, as Vicuna is open-source and requires fewer resources for training and deployment, the success of LogPrompt on Vicuna holds promising implications for building in-domain industrial applications.

\section{Discussion}\label{sec:discuss}
Based on our experimental findings, we identify several implications for LLM-based log analysis.

(1) Challenges in online anomaly detection. Without in-domain training data, the F1-score achieved by ChatGPT for anomaly detection is only 0.450, making it unsuitable for direct application in industry. The difficulty of online anomaly detection lies in the varied and flexible definitions of anomalies across different systems, which makes accurate detection through textual information in logs alone challenging. One potential solution is to incorporate general knowledge of existing software systems into LLMs.

(2) Significance of log interpretation. The positive feedback from practitioners regarding the generated explanations indicates the potential for improved user experiences in using log interpretation tools for practical log analysis. Interestingly, the results in Table \ref{tab:CoT_prompt} demonstrate that when LLMs are asked to generate both predictions and interpretations simultaneously, there is an additional improvement in the accuracy of analysis outcomes. We attribute this to the assumption that by demanding interpretations, an implicit CoT is created in the prompt, causing LLMs to "recall" the clean general CoT samples from the pre-training stage. Therefore, introducing domain-related CoT samples (\emph{e.g.}, log interpretations) in the pre-training stage of LLMs may lead to even greater improvements.

(3) Complexity alignment of prompt strategy. Certain strategies discussed in this paper, such as the CoT prompt for log parsing, have not been extensively explored for other tasks due to concerns about complexity. The mainly used prompt in log parsing, the self-prompt, already achieves good performance (F1-score of 0.797), which suggests that log parsing is a relatively easier task compared to anomaly detection (best F1-score of only 0.450). Therefore, implementing a more complex prompt like CoT is unnecessary for log parsing, as CoT is more suitable for tasks involving mathematical or logical reasoning, such as anomaly detection \cite{wei2022chain}.

\section{Threats to Validity}\label{sec:threat}
Our study faces several potential threats:

(1) Labelling errors in public datasets. Despite the use of publicly available datasets that contain logs from various domains, the manually labelled contents in the logs (\emph{i.e.}, log templates and anomaly tags) are still susceptible to errors, as reported in previous studies \cite{le2021log,huo2023semparser}. The sheer volume of raw logs makes it infeasible to identify and correct all errors, which may result in lower absolute performance in analysis tasks. For example, We have identified logs in the BGL dataset that are labeled as abnormal but appear to be semantically normal (\emph{e.g.}, the log message "exited normally with exit code" is marked as abnormal), which might explain the lower Recall score of LogPrompt in the BGL dataset. However, relatively speaking, LogPrompt still demonstrates remarkable performance, and the results of the ablation study indicate its efficacy.


(2) Black box nature of LLMs. The pre-training dataset used for LLMs consists primarily of textual data derived from web pages, books, and Wikipedia \cite{brown2020language,ouyang2022training,touvron2023llama}. Due to the black box nature of LLMs, it is possible that these LLMs, including BERT~\cite{kenton2019bert}, have been exposed to certain log fragments and subsequently acquired relevant concepts (\emph{e.g.}, from Wiki pages related to log analysis) during pre-training. However, both recent studies \cite{jiao2023chatgpt,peng2023towards,kocmi2023large} and our ablation studies (Table \ref{tab:CoT_prompt} and \ref{tab:varying_LLM}) emphasize that, without an appropriate prompt strategy, it is hard to elicit the domain knowledge stored within LLMs and further facilitate their log analysis capabilities, which may be the reason why LogPrompt outperforms other LLM-based approaches, such as LogPPT~\cite{le2023log} and LogStamp~\cite{tao2022logstamp}.

(3) Underestimated Baselines. To emulate the online scenario, we trained all baseline methods on a limited portion of each in-domain dataset (\emph{e.g.}, 10\%), which may have led to lower performance compared to the results reported in the original literature. Notably, most baselines are not specifically designed for online scenarios and require substantial training data (\emph{e.g.}, 90\%) to achieve optimal performance. Therefore, the underestimated baselines pose a fairness concern in the comparisons, particularly considering the extensive amount of data used to pre-train LLMs. 

(4) Insignificance of human evaluations. The limited number of sampled logs for the human evaluation in our study makes it difficult to achieve statistical significance in terms of usefulness and readability scores. Furthermore, it is uncertain whether the working experience of the human evaluators sufficiently qualifies them to judge the usefulness of the explanations generated for log analysis. However, scaling the sample size can be difficult due to the nature of human evaluations, and we have made efforts to ensure that all participants possess substantial experience (\emph{i.e.}, at least 10 years) in the field of software O\&M. Nevertheless, the positive comments and feedback from these practitioners revealed promising merits of LogPrompt.

\section{Related Work}\label{related_work}
\subsection{Log Analysis Tasks}
\subsubsection{Log Parsing}
Log parsing is a well-established technique for reducing the size of large log volumes by extracting meaningful templates from raw logs. This enables more effective analysis, such as anomaly detection, by simplifying the log data. Traditional approaches focus on the aggregation of lexically similar logs, such as cluster-based methods~\cite{zhu2019tools,fu2009execution}, heuristic methods~\cite{du2016spell,makanju2009clustering}, and tree structure mining methods~\cite{zhang2017syslog,he2017drain,chu2021prefix}.

More recently, there has been a growing recognition of the importance of variables in logs. Variables are no longer viewed as the parts in logs to be ``abandoned'', but the key to achieve deeper understanding of logs. LogParse \cite{meng2020logparse} and LogStamp~\cite{tao2022logstamp} use a word-level classifier to learn variable patterns and recognize variables directly from raw logs. Huo \emph{et al.} \cite{huo2023semparser} model both instance-level and message-level semantics for variables in logs, leading to improved downstream analysis. Li \emph{et al.} \cite{li2023did} manually derive eight general categories for variables in software logs and achieves high accuracy in variable classification.

In addition, LogPPT \cite{le2023log} incorporates prompt tuning techniques and employs a BERT-like language model for the purpose of log parsing. However, LogPPT's dependence on fine-tuning and predicting virtual prompt tokens using domain-specific logs constrains its adaptability and interpretive capabilities.

\subsubsection{Log Anomaly Detection}
Anomaly detection in logs can be performed using either forecast-based methods or classification-based methods. Classification-based methods require both normal and abnormal training logs, while forecast-based methods only require normal historical logs for training.

Examples of classification-based methods include LogRobust \cite{zhang2019robust}, which uses an LSTM model and a general-purpose word embedding as features, and Lu \emph{et al.} \cite{lu2018detecting}, who use a convolutional neural network (CNN) to capture sequential log features. Le \emph{et al.} \cite{le2021log} use a BERT encoder to directly recognize anomalies in raw logs without the need for log parsing. Forecast-based methods include DeepLog \cite{du2017deeplog}, which extracts normal log patterns and considers a new log anomalous if it violates these patterns, and LogAnomaly \cite{meng2019loganomaly}, which uses domain-specific synonyms and antonyms to match new logs with existing templates.

\subsection{LLMs in Software O\&M}
We also noticed the concurrent work by Le \emph{et al.} \cite{le2023evaluation}, who evaluate the effectiveness of ChatGPT for log parsing, with less emphasis on developing intricate prompts. Our research distinguishes itself from other LLM-based approaches and prior studies in log analysis in several ways. Firstly, we explore LLMs-based approaches for both log parsing and anomaly detection tasks, unlike the single supported task of log parsing in \cite{le2023log,le2023evaluation}. Secondly, we provide scientific descriptions of prompt construction and validate the effectiveness of three advanced prompt strategies in log analysis, in contrast to the simple prompt strategies employed in \cite{le2023log,le2023evaluation}. Thirdly, this is the first work to address the challenge of limited interpretability in log analysis, which is an area that has been overlooked in previous research. To evaluate interpretability, we conduct a rigorous human evaluation guided by detailed interpretability evaluation criteria. Lastly, our proposed approach is highly adaptive and does not require in-domain training, unlike existing work that still relies on resources for an in-domain training stage \cite{le2023log}.

Although this is among the very first studies to apply LLMs to log analysis, there have been several studies exploring the use of LLMs in software O\&M. Ahmed \emph{et al.} \cite{ahmed2023recommending} fine-tune GPT-3.x models to recommend root causes and mitigation steps for cloud incidents and propose a scoring schema for human evaluation of the generated recommendations. Somashekar \emph{et al.} \cite{somashekar2023enhancing} propose adapting language models to the task-specific domain through training on system configuration files to enhance the configuration tuning pipeline for distributed applications. In contrast to these studies, which fine-tune LLMs with large amounts of training data in a specific domain, our work explores general prompt strategies for LLMs to perform log analysis that can be applied to any domain.

\section{Conclusion}\label{sec:conclusion}
In this study, our objectives transcend the simple application of the latest advancements in LLMs within log analysis. Instead, we aim to offer a scientific and comprehensive engineering methodology to accurately discern how to optimize prompts to enhance the performance of LLMs within log analysis, and, crucially, to comprehend the potential areas where LLMs can aid engineers in software O\&M and program comprehension. Although the advanced measures of LogPrompt in online scenarios augment adaptability in log analysis by superseding the labor-intensive in-domain training process, we believe that LogPrompt's primary and unique advantage resides in facilitating program comprehension by furnishing justifications alongside predictions. In our expert study, practitioners provided positive anticipations on the application of LogPrompt in real-world scenarios, including event comprehension in unfamiliar domains, composition of analysis reports, and ruling out false alarms. In addition, we constructed a systematic and intricate criteria to assess the usefulness and readability of these generated explanations in log analysis, which may inspire future research aimed at improving interpretability scores and facilitate industrial applications. Lastly, we substantiate that LogPrompt is compatible with open-source and smaller-scale LLMs, thereby validating its feasibility for actual deployment.



Future research directions include advancing the ability of LLMs through domain adaptation techniques, validating LogPrompt on more LLMs and expanding prompt strategies to more tasks.  

\clearpage
\bibliographystyle{ACM-Reference-Format}
\bibliography{mybib}


\end{document}